# An Overview of the LIGO Control and Data Acquisition System[*]

R. Bork, R. Abbott, D. Barker, J. Heefner, LIGO Laboratory,
California Institute of Technology, Pasadena, CA 91125, USA

Abstract

The LIGO Control and Data system (CDS) features a tightly coupled and highly integrated control and data acquisition system. Control of the interferometers requires many Multiple Input Multiple Output (MIMO) control loops closed both locally and across the 4-kilometer interferometer arm lengths. In addition to providing the closed loop control, the control systems front end processors act as Data Collection Units (DCU) for the data acquisition system. Data collected by these front ends and the data acquisition system must be collected and time stamped to an accuracy of 1 microsecond and made available to on-line analysis tools such as the Global Diagnostics System (GDS)[1]. Data is also sent to the LIGO Data Analysis System (LDAS)[2] for long-term storage and off-line analysis. Data rates exceed 5 Mbytes per second per interferometer continuous. Connection between the various front end processors and the data acquisition system is achieved using fiber optic reflective memory networks. Both controls and data acquisition systems use VME hardware and VxWorks® operating systems. This paper will present an overview of the LIGO CDS and discuss key aspects of its design.

## 1 INTRODUCTION

The LIGO interferometers located in Hanford, Washington and Livingston, Louisiana are Michelson laser interferometers enhanced by multiple coupled optical resonators. These coupled optical resonators are 4-kilometer long Fabry-Perot cavities placed in each arm of the interferometer. The mirrors that form the cavities are suspended from a single loop of wire mounted inside suspension cages that are mounted on seismically isolated optical platforms within the LIGO vacuum system.

The Control and Data System (CDS) for these interferometers must provide many features. Among the requirements are:

- Provide for monitoring and control of the vacuum systems in which the interferometers reside.
- Provide for continuous data acquisition at rates to 5Mbyte/sec per interferometer.
- Provide for interferometer diagnostics to monitor interferometer systems and measure interferometer performance.
- Provide for local damping of all suspended optics.
- Provide for pitch and yaw (alignment) control of all suspended optics.
- Provide arm length control for four independent degrees of freedom. Each of these lengths must be controlled to an integral number of half-wavelengths of the laser light ($\lambda = 1.06\mu m$) with high accuracy, ranging from nm to less than 0.1pm.
- Provide the integration mechanisms to bring the system from a group of individually damped mirrors to a locked interferometer [4] with a strain sensitivity of $10^{-21}/\sqrt{Hz}$.

## 2 CONTROL SYSTEM DESIGN

### 2.1 Overview

The following figure shows an overview of the interferometer and its control system.

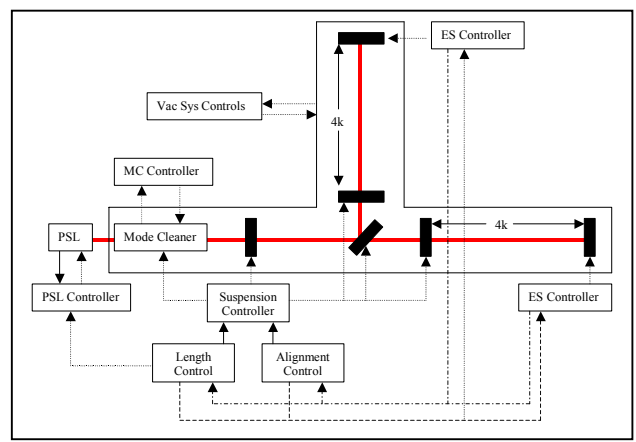

The interferometer itself consists of a pre-stabilized laser (PSL), a mode cleaner with seven suspended optics, and five large core optics. The core optics consist of a recycling mirror, beam splitter, two input test masses and two end test masses.

The LIGO laser is designed as a Non-Planar Ring Oscillator (NPRO) in a Master Oscillator Power

---

[*] Work supported by the National Science Foundation under Grant PHY-920038.

Amplifier (MOPA) configuration. The laser was designed and built commercially to LIGO specifications. The unit operates at 10W, with an output wavelength of 1064 nm. To the laser system, LIGO CDS adds in-house designed frequency and power stabilization.

At the output of the PSL, a mode cleaner, mode matching telescopes and steering mirrors are provided to direct the beam into the interferometer. For these optics, as well as the core optics, CDS provides for local damping, alignment and length control.

## 2.2 Building Blocks

In general, the CDS front end controls are based on VME systems. To the extent possible, Commercial, Off-The-Shelf (COTS) VME modules are employed. Custom VME module designs are primarily limited to timing system interfaces. A basic block diagram of a typical VME subsystem is shown in the following figure.

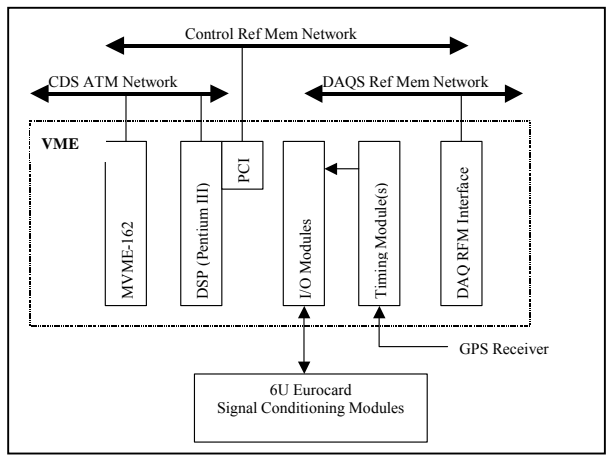

Processors in the VME systems are of two types: 1) Motorola MVME162, for slow (<10Hz) control and operator communication applications and, 2) PentiumIII processors, for real-time Digital Signal Processing (DSP) applications (to 16KHz). In the case of the high bandwidth servo applications, both an MVME-162 and DSP will reside in the same VME crate. The MVME-162 performs the task of interfacing information between operator stations, via the CDS ATM network, and the DSP, via the VME backplane. The DSP only connects to the CDS ATM network for purposes of downloading software.

VME I/O modules are comprised primarily of Analog to Digital Converters (ADC), Digital to Analog Converters (DAC) and binary input/output modules. All ADC and DAC are simultaneous sampling 16 bit devices, with eight individual ADC and DAC per VME module.

Timing clocks for controls and data acquisition are derived from the Global Positioning System (GPS). A commercial GPS receiver was developed for LIGO that provides a $2^{22}$Hz clock output, phase locked to the GPS 1Hz clock. LIGO custom timing boards to clock and synchronize all of the LIGO control systems and data acquisition system distribute the 1Hz and $2^{22}$Hz clocks.

For interfacing to the various interferometer sensors and actuators, a large number of custom modules were developed in 6U Eurocard format for installation in 19" rack mount Eurocard chassis. Types of modules designed in-house include whitening/dewhitening filters, anti-aliasing/imaging filters, and I&Q demodulators. Several closed loop controls are also performed in analog circuitry or analog circuitry combined with digital controls. Examples include the pre-stabilized laser and mode cleaner controls.

For communications between the various processors involved in control and data acquisition, three networks are provided:

- CDS Asynchronous Transfer Mode (ATM) network. This network consists of an ATM switch with direct connection to operator stations and servers and uplinks to Ethernet switches at various locations for connection to the VME processors. This network is used for downloading code to VME processors and communicating information between operator stations and the VME processors.
- Data Acquisition Network. This network is based on 1Gbit/sec reflected memory, with up to 4Mbyte of memory per node.
- Control Reflected Memory (RFM) network. This network is based on 240Mbit/sec, 64Mbyte per node, reflected memory. This network is used for real-time communication between DSPs. The RFM interface modules directly connect to the DSP as PCI Mezzanine Cards (PMC).

Backend computers and servers are typically Sun Microsystems workstations and servers. These are employed as operator stations, system servers, file servers, and for interferometer diagnostic data analysis.

Software development for CDS is based on operational requirements. For all real-time DSP applications, code is developed in C and runs on the vxWorks® operating system on VME based PentiumIII processors.

For general network communications and slow controls, the Experimental Physics and Industrial Control System (EPICS), first developed at Los Alamos National Laboratory, is used. The primary features of EPICS used in the LIGO CDS are channel access, used to communicate information between operators and the VME processors, the graphical user interface features, for development of operator displays, and the back up and restore tools.

For IFO diagnostic analysis and data display, code is developed in C++ for a ROOT based system originally developed at CERN.

## 2.2 Vacuum Controls

The vacuum chambers, vacuum tubes, vacuum pumping and all associated sensors and actuators were designed, built and installed under contract by commercial vendors. LIGO provided the control and monitoring systems. This system was designed as a stand-alone system, with three 19" racks of equipment in the corner station and one in each mid and end station. The controls are VME based, using MVME-162 processor boards running EPICS software.

Actual monitoring and control is provided by software written using the EPICS State Notation Language (SNL) feature. For each type of device to be controlled and/or monitored, a generic EPICS database and SNL program was developed and tested, much like a C++ object. A PERL script was then used to generate the unique instances required by each VME processor in the system. A second script generates a startup file based on the contents of this directory. The file is loaded by the VME processor on power up or reset.

## 2.3 Laser Controls

The commercial laser system is refined in both amplitude and frequency stability by dedicated feedback controls. The entire system is referred to as the Pre-Stabilized Laser (PSL). An overview of the PSL is shown in the following figure.

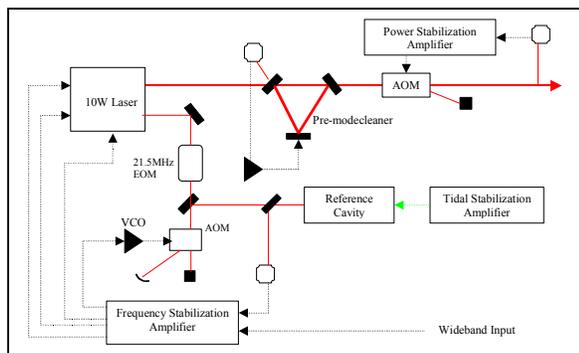

The frequency stabilization servo must initially reduce the intrinsic laser frequency noise by a factor of 1000. To meet the ultimate frequency noise requirements of LIGO, additional frequency stabilization is implemented by a "wideband" correction signal derived from the interferometer driving a dedicated PSL frequency adjusting input.

The amplitude stabilization servo was implemented to provide up to 80 dB of suppression of amplitude jitter inherent to the laser output. Both the frequency and amplitude stabilization are fully controllable through remote user interfaces. Additional amplitude stabilization and isolation of the fundamental optical mode is provided by the pre-modecleaner servo, which functions analogously to a tracking band-pass optical filter.

## 2.4 Interferometer Controls

The interferometer controls that provide for the local damping of optics (Digital Suspension Controls (DSC))[4], Alignment Sensing and Control (ASC) and interferometer Length Sensing and Control (LSC) are provided by MIMO digital servo controllers. All of these control subsystems share common design features.

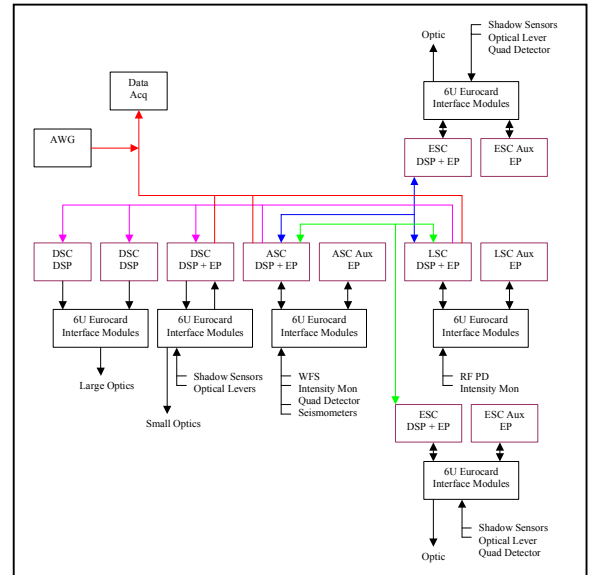

Each of these digital control systems (DCS) consists of a VME crate with two CPUs, one DSP and one MVME-162 processor running EPICS (EP). The EPICS CPU has a database for communicating information to/from operators, with SNL code providing communications between the EPICS database records and the DSP. Essentially, this SNL code reads/writes local memory mapped to the VME backplane. The DSP updates this memory at 16Hz. The DSP closed loop control code runs at either 16384Hz or 2048Hz, dependent on the servo bandwidth requirements.

While the software for each DCS requires some custom software particular to the application, a fairly standard pattern for control is consistent in all units and code libraries have been written for these standard features. The standard software design is shown in the following figure.

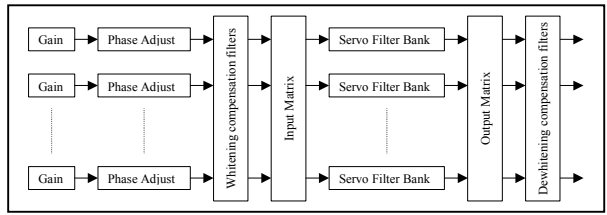

As required, each input sensor signal is provided with a gain and phase shift block, settable by the operator. Some controllers require switchable hardware whitening filters. As these are switched in and out, software filters must also be switched to compensate for the hardware and maintain loop stability.

The input signals are then passed to an input matrix. The matrix parameters can be set by operators, or, as in the case of the LSC, this matrix can be automatically adjusted based on input parameters.

The actual servo calculation filter banks follow the input matrix. These filter banks contain as many as ten filters each. An output matrix and dewhitening compensation filters follow the filter bank. In some cases, the output matrix is actually a matrix of filters.

All of the digital filters in these systems are based on second order section (SOS) IIR filters. A filter bank consists of:

- A primary input, with gain and offset adjusts.
- An auxiliary input for insertion of excitation signals from the GDS.
- Up to 10 filter sections, each of which may be up to an $8^{th}$ order filter.
- An output inverter stage.
- An output enable/disable.

Initially, the filter coefficients were evolved from modeling of system requirements and then placed in code header files. As is typical, during interferometer commissioning it was found that the theoretical models and real world did not exactly match and the changing of filter coefficients became common. Therefore, the systems have evolved to writing the filter coefficients to separate files, which are user modifiable, and can be uploaded to the DCS. This is done in such a way that filters can be changed "on the fly", i.e. no task restart or CPU reboot is required.

While the DSC, ASC and LSC are individual subsystems of CDS, they must communicate with each other in real-time. Certain sensors are connected to one subsystem, but are also required by another subsystem and control outputs of the ASC and LSC must be fed to the DSC for actual output to the optics. The sensors and control actuators may be up to 4-kilometers away from the DSP performing the servo control functions. To facilitate these communications, the CDS RFM control network is used. Each interferometer control system has three of these networks: one for communication within the corner station and one for communication with each end station. The use of three networks was required to minimize and balance the servo loop phase delay, as the speed to light to the end stations adds 13μsec.

## 3 DATA ACQUISITION

The LIGO Data Acquisition System (DAQS) provides several key functions:

- DAQS must acquire data from digital controllers and various analog signals from LIGO sensors and record this data to disk for consumption by the LIGO LDAS system for analysis.
- DAQS must broadcast acquired data to LIGO GDS computers for interferometer performance analysis and reduced data set writing.
- DAQS provides the network connections between the GDS Arbitrary Waveform Generator (AWG) and the digital controllers.
- DAQS provides for data transmission to operator stations and graphical tools to display the data.

An outline of the LIGO DAQS is shown in the following figure.

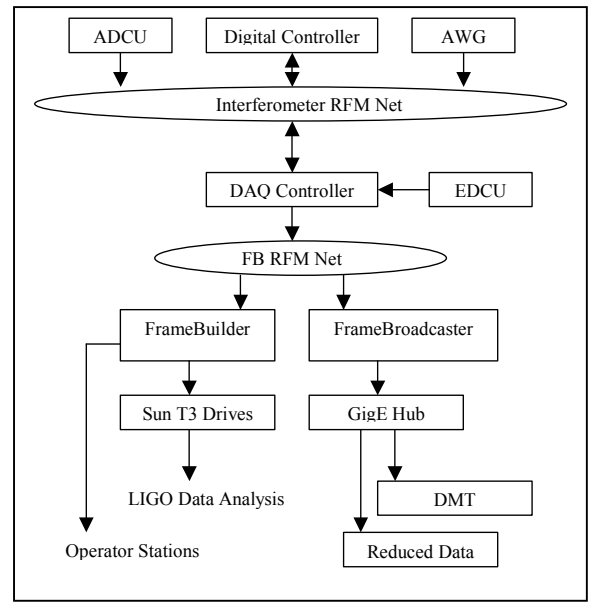

Input to the DAQS can be either as direct analog input signals, via the Analog Data Collection Units (ADCU) or as digital signals from the various LIGO digital servo controllers. These units are interconnected to each other and the DAQ controller via a 1Gbit/sec Reflected Memory (RFM) network. Each processor on the network is assigned a defined block for transferring its data and communicating control and status information with the DAQS controller.

ADCUs are VME crates with a processor, RFM network module, timing system connections, and up to five Analog to Digital Converter (ADC) modules. The ADC modules used in the system contain 32 individual, 24-bit sigma delta ADC with programmable gains. Signals are connected to the ADCU via an anti-aliasing chassis, which contains plug in filters appropriate for the data acquisition rate of a particular channel. Acquisition rates are selectable from 256Hz up to 16384Hz for each channel in steps of $2^n$.

The DAQS RFM network also provides a connection for the Arbitrary Waveform Generator (AWG), provided as part of GDS. The AWG uses the DAQS RFM to inject excitation signals into defined points of the digital servo controllers or can directly produce analog outputs via its own Digital to Analog Converters (DAC).

All EPICS channels are available to be collected by the system at rates to 16Hz. As part of the DAQ controller VME crate, a second CPU, the EPICS Data Collection Unit (EDCU), collects requested EPICS channels and stores them in local memory. This memory is then read across the VME backplane by the DAQ controller for insertion into the main DAQS data stream.

The DAQ controller provides for configuration and monitoring of the various DAQS components and reformats data received from the various Data Collection Units (DCU) and transfers it on to the FrameBuilders. Configuration of the system is read from a text file on startup or operator request. This file contains information on all channels to be acquired, including connection points, acquisition rates, connection type, data storage requirements and conversion parameters to engineering units. Acquisition channels may be classified as either permanent connections, which are acquired and stored continuously, or as test points, which are only introduced into the system on operator request and are not stored.

The DAQ controller reads data from the interferometer RFM network and from the EDCU each 1/16 of a second. It reformats the data and then transfers it on to the FrameBuilder and FrameBroadcaster via a second RFM network.

The primary function of the FrameBuilder is to format the data into LIGO standard data frame format and then store that data to disk. The frame format is a standard adopted by the gravitational wave community for the sharing of data. A frame contains descriptor information on all of the data contained within a frame and data for all channels acquired in a given time period. At present, LIGO is producing frames that contain 16 seconds of data. Data written to disk are kept on line for up to three weeks before being archived by the LDAS.

In addition to these frames, which contain all data in raw format, the FrameBuilder also produces what are called trend frames. These are files that contain statistical information about the data channels, at one second, 1 minute and 1 hour intervals.

For delivery of data to operator workstations, the FrameBuilder also provides a Network Data Server (NDS). The NDS provides data on request via a CDS defined lightweight protocol. NDS is used by the DAQS data display software and GDS tools to retrieve information from the DAQS. Data is made available near real-time, at 16Hz update rates, or from LIGO data frames or trend files on disk.

The FrameBroadcaster produces one second data frames. However, rather than storing the data to disk, this unit broadcasts data in frame format over a gigabit Ethernet. This information is then received by several GDS interferometer diagnostic computers, which use the data to characterize various aspects of interferometer performance. The data broadcast is also received by the Reduced Data Set (RDS) writer. This unit records subsets of data and also has the ability to store test point data, which is not recorded by the main FrameBuilder. The RDS is typically used to record short stretches (hours) of data for off-line analysis in support of interferometer commissioning.